\documentstyle[preprint,eqsecnum,aps]{revtex}
\begin{document}
\draft
\preprint{twogat05.latex--submitted to Phys. Rev. A (AT5101); condmat/9407022}
\title{Two-bit gates are universal for quantum computation\\
}
\author{David P. DiVincenzo
}
\medskip

\address{
IBM Research Division\\
Thomas J. Watson Research Center\\
P. O. Box 218\\
Yorktown Heights, NY 10598 USA\\
}
\date{Received June 24, 1994}
\maketitle
\begin{abstract}
A proof is given,
which relies on the commutator algebra of the unitary Lie groups,
that quantum gates operating on just two bits at a time are
sufficient to construct a general quantum circuit.  The best previous
result had shown
the universality of three-bit gates, by analogy to the universality
of the Toffoli three-bit gate of classical reversible computing.
Two-bit quantum gates may be implemented by magnetic resonance operations
applied to a pair of electronic or nuclear spins.  A ``gearbox quantum
computer" proposed here, based on the principles of atomic force microscopy,
would permit the operation of such two-bit
gates in a physical system with very long phase breaking (i.e., quantum
phase coherence) times.  Simpler versions of the gearbox computer
could be used to do experiments on Einstein-Podolsky-Rosen states
and related entangled quantum states.
\end{abstract}

\narrowtext

\section{Introduction}
\label{sec:level0}

Eventually computational devices will stop getting smaller
and faster unless new
physical principles of operation are discovered.  Undoubtedly the physical
principles
of quantum mechanics will become increasingly important if
these devices are ever to operate at the atomic level.
The basic idea that a useful computer might be constructed which operates
according to the principle of unitary time evolution, the cornerstone
of the quantum theory, was first put forward by Benioff\cite{Beni}.
Since then,
there has been a steady stream of work on demonstrating how ``quantum
gates" can be put together into ``quantum circuits" which perform any
unitary time evolution, and on how such quantum gates could be realized
by the electromagnetic pulsing of solid-state spin systems.

While these developments have been of theoretical interest, the importance
of these investigations has been heightened by some recent seminal
mathematical results concerning the potential power of quantum computing.
Deutsch and Josza\cite{Deu92}, with some crucial clarifications from
Bernstein and Vazirani\cite{BV},
introduced the subject of quantum complexity theory,
and pointed out the possibility that quantum machines may be more
efficient in performing certain computations than any classical computer.
Very recently Shor\cite{Shor},
following on the work of Simon\cite{Simon}, has proved an extremely
exciting result: he has shown that on a quantum computer, prime factoring
can be performed in ``polynomial" time, that is, $t\propto k^p$, where
$k$ is the number of
bits in the number to be factored and $p$ is a constant.
By contrast, this problem is believed to take
$e^{ck^{\frac{1}{3}}}$ time on a classical computer
($c$ is another constant).
As the difficulty of prime factoring
is of paramount importance in  the functioning of certain popular data
encryption schemes, the absolute desirability of performing quantum
computation, and the interest in understanding how a genuine physical
realization might be achieved, has increased sharply.

In this paper, then, I take up some specific problems which will need to
be addressed in order to make quantum computing a reality.  I begin with some
discussion which emphasizes the stringent requirements in quantum
computing for the {\em physical isolation} of the computer from outside
influences; this point has been made  in much of the previous work, but
I wish to emphasize it as probably the most difficult design requirement.
Motivated by this, I introduce a computing machine, a ``quantum gearbox",
which arranges for individual spins to be in very well-isolated environments
except during the moment that pairs of spins pass through logic gates.
I will point out
here that in the ``quantum gearbox", and in other physical implementations
which have been proposed for quantum circuits as well, it is extremely
difficult to imagine a physical implementation of a three-bit quantum
gate, that is, a gate in which three bits (i.e., spins) interact
simultaneously.  It is much easier (although not necessarily easy) to
imagine a machine in which spins interact two at a time (this is explicit
in the quantum gearbox).

This situation motivates the main results of the paper, on the universality
of two-bit computation.  It has previously been proved that three-bit gates
are sufficient to build any arbitrary quantum network, and no other
workers have investigated whether this result can be improved upon.
Here, using the techniques of Lie group theory, I prove the desired new result,
that two-bit gates suffice to generate any arbitrary quantum network,
i.e., any arbitrary unitary transformation.  The proof provides an
explicit realization of three-bit operations in terms of sequences of
two-bit gates, although it remains to be seen whether this may form the basis
of
a practical, efficient method of designing quantum circuits.  Substantial
progress has already been made in devising explicit two-bit-gate
realizations of some of the key steps in the Shor factoring
procedure\cite{Copper}.

\section{Building a quantum computer}
\label{sec:two}

\subsection{Why making a quantum computer is very, very difficult}
\label{sec:vvd}

As Shor's work shows,
making a quantum computer would have decided technological
and economic consequences.  Why will neither IBM, nor Dell, nor anyone,
be marketing one before the end of the century?  The laws of physics give us
confidence that the world does indeed evolve by unitary time evolution (i.e.,
according to an S-matrix--see Sec. \ref{sec:level2}).
The problem is that to make a quantum computer,
we insist that a {\em particular
subset} of the world undergo unitary evolution; this is what is very, very
hard.  A sub-block of a unitary matrix is almost never itself unitary--it would
be so only if the
matrix were block-diagonal.  An S-matrix is only block diagonal if
the different
subsystems are not interacting; but in the physical world, degrees of freedom
are usually interacting with many other degrees of freedom.  The understanding
of this point is crucial for the explanation of why classical mechanics in
the macroscopic world emerges out of the microscopic operation of quantum
mechanics.

This discussion makes clear why a transistor, or any conventional computer
element, cannot perform quantum computation.  The computational state of
the system, the ``0"'s and ``1"'s entering and leaving
the gate, is only one degree of freedom out
of the countlessly many microscopic degrees of freedom of the device (e.g.,
the elastic vibrations of the device, the excitations
of its conduction electrons).
In general, all of these degrees of freedom interact strongly with one another
and with the computational state of the device.
Even worse, in fact, is that the computational state is often a collective
property of this myriad of microscopic states.  Such a situation makes even
approximate unitary evolution impossible.

The kind of subsystem-isolation which quantum computation requires will
probably
only be achievable if the computer elements are themselves of atomic or
near-atomic dimensions, where the computational state is the quantum state
of a single atom.  Even in this realm, quantum computation is under substantial
constraint: if this computation state is arranged to interact weakly with the
rest of the world, then for short times its evolution will be unitary, but
eventually even weak interactions will cause significant departures
from unitarity.
Such systems have a characteristic time for loss of unitarity, which is
known in the field of mesoscopic physics as the ``dephasing time"
$t_\phi$\cite{Wash}.
$t_\phi$ has been measured in various microscopic and mesoscopic physical
systems, and it is often extremely short.  For example, for the state of an
electron traversing a gold wire at temperatures less than 1K,
$t_\phi$ is of order $10^{-13}$ seconds.  (This time is still long enough
for interesting ``phase coherence" effects to be seen, such as Aharonov-Bohm
oscillations\cite{Wash}.)
The state of an electron's spin\cite{Band}
(i.e., the state of the
electron's magnetic moment) is more stable, but an upper bound for its
dephasing time, recently measured in a salt containing
paramagnetic Eu ions\cite{Eu}, is $10^{-3}$ seconds.
Since the program on a quantum computer would have to be finished running
in a time less than $t_\phi$ (probably a great deal less), we see that there
are severe limits on the kind of quantum computation which these physical
systems can perform.
I believe that other microscopic systems which have been discussed for
quantum computation, for example the ``Notre-Dame logic gate"\cite{Poro}
(operating
by the hopping of electrons from one quantum dot to another) or the ``atom
switch"\cite{Eig}
(operating by the hopping of a single atom from one site on a crystal
surface to another) are similarly problematic;
although I know of no measurements of $t_\phi$ in
these cases, I expect that it is similarly short.
Even these systems will be ``too classical"\cite{foot1}.

Although still far from easy, I believe that one of the other microscopic
systems which has been discussed in this context has a somewhat greater
promise of having the necessary quantum coherence for some kind of quantum
computation to be possible:  this is the nuclear spin.  The spin of the
nucleus produces a much smaller magnetic moment than that of the electron
(650
times smaller for the proton), so its dipolar magnetic interactions with the
rest of the world are much weaker; also, it does not have the same strong
exchange
interactions arising from the Pauli exclusion principle which electrons do.
For this reason, nuclei have $t_\phi$'s which
can, under favorable circumstances, be orders of magnitude longer than for
electronic spins, or any other quantum degrees of freedom.  The value of
this $t_\phi$ has been estimated, given ideal assumptions about the
electromagnetic environment in a Nuclear Magnetic Resonance apparatus,
to be $10^{10}$ seconds
(3 years)\cite{Abr,Lloyd2}.  This number will certainly be much smaller
in reality and will
depend on many details of the solid-state environment of the spins,
which I will discuss in a moment.

\subsection{Gearbox quantum computer}
\label{sec:gearb}

In Fig.~\ref{qcomp01}$\ $ I propose the principal working element
of a quantum computer.
It is meant to be more thought-provoking than real: I don't suggest that
experimentalists try immediately to go out to build this device, but I do
hope that it provides a springboard for productive thought on what really
needs to be
done to perform quantum computation.  Of course, others before
me\cite{Lloyd,Obe}
have made proposals for, and explored the feasibility of\cite{Berm},
``potentially realizable
quantum computers"; the one I now present exploits somewhat different physical
principles than the previous proposals, and hopefully provides ideas for how
some of the monstrous
obstacles (like the ones discussed in Sec.~\ref{sec:vvd})
could be overcome.

In the ``gearbox quantum computer" shown, the two meshed gears operate
classically, turning in synchrony.  The protons, carrying the spins which
will evolve quantum mechanically,
are firmly
attached to the end of the tips of the left-hand gear, and to the base
of the grooves of the right-hand gear.
By making the gear elements very ``quiet" magnetically and
electronically, it may be hoped that a very long dephasing time
for the spins may be achieved.  This quietness may be obtained by using
diamagnetic, insulating materials, containing nuclei with mostly no
nuclear spins.  A gear made from a pure undoped crystal of
$\ ^{28}{\rm Si}$ (92\% natural abundance)
could well be optimal.

The teeth of the left-hand gear are shown in the
shape of atomic-force microscope (AFM) tips, suggestive of the fact that
atomic spatial resolution will be necessary in the meshing of the two
gears, in order that the two spins may be brought into atomic contact.
The gears are shown with 16 and 15 teeth respectively; by making these
numbers relatively prime, it is assured that each pair of spins on
either gear may be brought into contact by turning the gearbox.
The atomic contact between spins is necessary in order for quantum
logic gates, unitary transformations of pairs of bits, to be executed.
When the spins are not in contact, the Hamiltonian
$H_{spin}$ of the spins is
zero, and no time evolution occurs.  During the time that the spins
are in contact, $H_{spin}(t)$ will be non-zero, inducing, according
to the well-known laws of quantum mechanics\cite{Neg},
the unitary transformation
\begin{equation}
U=Te^{-\frac{i}{\hbar}\int H_{spin}(t)dt},\label{tord}
\end{equation}
where $T$ indicates a time-ordered product.  We can write out the
Hamiltonian a little further as $H_{spin}=H_{dipole}+H_{ext}$, where
$H_{dipole}$ is the magnetic dipole interaction between the two
nuclei, and $H_{ext}$ is an external Hamiltonian which may be
applied just to the region where the spins are interacting; this
may consist of some combination of static and a.c. magnetic fields.
It is expected that, as in the work of Lloyd\cite{Lloyd},
it is possible with suitable external fields to induce any arbitrary
two-spin unitary operation in Eq. (\ref{tord}), although I have not
worked out a detailed protocol for this.  In the next section
I will discuss the adequacy of these two-bit gates for general
quantum computation.

A few more remarks about the ``quantum gearbox" are in order.
It is obvious that if more bits are needed, for I/O, memory, etc.,
they can be added simply by adding more gears to the system.
Note that since one of the
simplest unitary operations is a swap, the state of any spin may
be propagated arbitrarily far along an array of gears.  Somewhere in
the gear system will be located the ``output" device, which will require
considerable technological ingenuity:  This device must sense the
state of a single spin and make that information available to the
rest of the world.  This operation can be done only at the end of
the quantum computation, since it involves strong interaction of
the quantum computer with other degrees of freedom, destroying
the unitary evolution.  There is presently no magnetometry of
sufficient sensitivity to sense the state of a single proton spin;
however, the mechanical detection of magnetic resonance within
magnetic force microscopy will, according to Rugar and coworkers\cite{Rugar},
be able to perform such detection in the foreseeable future.

There is another design requirement for a quantum computer whose satisfaction
requires some ingenuity in
the quantum gearbox.  Quantum computations under consideration
now\cite{Simon,Shor}
require that the spins be an initial simple state,
e.g., all up.  But because of the low energy scales involved, an
assemblage of nuclear spins
at any reasonable temperature
will typically have a random state which is drawn from a Boltzmann
distribution.  Nevertheless, it would be possible to exploit some
of the techniques of magnetic resonance to prepare an initially
polarized state.  One way of doing this would be to prepare a
gear with attached single {\em electron} spins.  Because of their
much greater magnetic moment, these can be put in a polarized state
at a reasonable temperature.  When they are meshed into contact
with the nuclear spins, one can use one of the
known techniques (the
Overhauser effect, coherence transfer\cite{NMR}), for transferring
the electronic spin polarization to the nuclei.  This process
need not be phase-coherent, because it would preceded the start
of the quantum computation.

I wish to close this section with a few remarks on the strengths
and weaknesses of the quantum gearbox relative to another ``potentially
realizable quantum computer" discussed recently by Lloyd\cite{Lloyd},
in which the quantum spins are embedded in a polymer chain or a crystal
lattice.  One obvious advantage of the polymer computer is that the
interaction Hamiltonian between the spins is more controlled, since
it depends only on the local environment in the crystal.  In the
gearbox computer, atomic-scale vibrations and misalignments might
be very difficult to control and quite deleterious to the operation.
Another advantage of the polymer computer is that the unitary operations
can go on in parallel, although the ability to address specific pairs
of spins is lost.  In the polymer computer, not just two-bit but
also three-bit local operations can be executed (but this is not
a crucial advantage--see the next section).  The polymer computer
has the disadvantage that the interaction Hamiltonian between the
spins can never be turned off.  This does not necessarily lead to
insuperable problems, but it makes the control of the phase of
the quantum state particularly ungainly.  Finally, there is a concern
that it may be difficult to make a magnetic polymer or a magnetic
crystal sufficiently ``quiet", i.e., sufficient immune from interaction
with other, non-computational degrees of freedom.

\section{Demonstration of two-element gates\protect \\
for universal computation}
\label{sec:level1}

Previous studies, to be reviewed momentarily, have shown how to perform
any arbitrary unitary operation by composing a sequence of three-bit
operations.  This is very inconvenient from the point of view of the
gearbox computer; it is exceedingly difficult to imagine a mechanical
device which could bring three spins together simultaneously.  However,
I will prove a new result which resolves this difficulty: even two-bit
operations alone suffice to give universal quantum computation.

\subsection{Background: What Deutsch proved}
\label{sec:level2}

Deutsch \cite{Deutsch89} has already shown how to obtain a universal
quantum computation, defined as an arbitrary unitary transformation
on a discrete Hilbert space spanned by the set of all states of a
collection of bits.
He did this by a simple and elegant generalization of the known specifications
for building a reversible {\em classical} network.
There exists a close connection between
classical reversible computation and quantum computation, since all
unitary quantum operations are necessarily reversible; therefore, reversible
computing is a subset of quantum computing.  Toffoli\cite{Toffoli80}
showed how the AND and XOR gates necessary for conventional universal
computation may be implemented reversibly; conventional AND and XOR
gates are not reversible, if for no other reason that a reversible gate
must have the same number of output as input bits.  He showed that XOR
could be implemented reversibly with a two-bit gate in which one output
bit returns the conventional XOR $a_1\oplus a_2$ ($a_1$ and $a_2$
are the binary
values of the two input bits), while the other output bit returns the
original value of $a_1$ (or $a_2$).  To implement AND reversibly, a three-bit
gate is required in which $a_1$ and $a_2$ are passed through unchanged, while
the third bit is XORed with the AND of the first two, returning $a_1\cdot a_2
\oplus a_3$.  Indeed, since this three-bit gate comprises both the XOR and the
AND functions, it can be considered to be {\em the} universal reversible
computation gate, and it has come to be known as the Toffoli gate {\bf T}.

Given this background, the generalization by Deutsch to the quantum
problem is simple and appealing.  Following logically from the structure
of quantum mechanics, Deutsch generalized the posited operation of a
three-bit gate, from one which performs transformations (permutations,
actually, in the reversible case) on the $8=2^3$ possible states of three
bits, to one which performs unitary transformations within the
$2^3$-dimensional complex vector space (the ``Hilbert space") spanned
by the states of the three bits.  Deutsch proved that all unitary
transformations could be obtained from one operating upon three bits,
that one being a natural generalization of the Toffoli operation.
This result has been used in subsequent studies\cite{Yao} to understand the
complexity of quantum circuits using three-bit gates.

Deutsch's universal gate {\bf Q} has the S-matrix\cite{foot2}
\begin{equation}
({\bf S_Q})^{a_1a_2a_3}_{a'_1a'_2a'_3}\ =\ \delta^{a_1}_{a'_1}
\delta^{a_2}_{a'_2}[(1-a_1\cdot a_2)\delta^{a_3}_{a'_3}+ia_1\cdot a_2
e^{-\frac{1}{2}i\pi\alpha}(S^\alpha_N)^{a_3}_{a'_3}]\label{one}.
\end{equation}
``S-matrix" is the quantum mechanical jargon for the unitary transformation
executed (in the course of a given length of time, say) upon the Hilbert space.
Here the primed variables denote the binary states of the three output bits
(as in Fig.~\ref{qcomp02}),
$\alpha$ is a fixed arbitrary irrational number, and $S^\alpha_N$ is an
elementary one-bit transformation specified by the $2\times 2$ unitary matrix
\begin{equation}
S^\alpha_N\ =\ \frac{1}{2}\left(\begin{array}{cc}1+e^{i\pi\alpha},&
1-e^{i\pi\alpha}\\1-e^{i\pi\alpha},&1+e^{i\pi\alpha}\end{array}\right)
\label{two},
\end{equation}
to which Deutsch gives the
picturesque appellation ``the $\alpha^{th}$ power of not".
It is noted that
when $\alpha=1$,
except for a phase factor
Eq. (\ref{one}) is the S-matrix
of the classical Toffoli gate:
\begin{equation}
({\bf S_T})^{a_1a_2a_3}_{a'_1a'_2a'_3}\ =\ \delta^{a_1}_{a'_1}
\delta^{a_2}_{a'_2}[(1-a_1\cdot a_2)\delta^{a_3}_{a'_3}+a_1\cdot a_2
(S^{\alpha=1}_N)^{a_3}_{a'_3}]\ =\ \delta^{a_1}_{a'_1}
\delta^{a_2}_{a'_2}\delta^{(a_3\oplus a_1\cdot a_2)}_{a'_3}\label{three}.
\end{equation}

\subsection{Proof that {\bf Q} can be realized by two-bit gates}
\label{sec:level3}

In this subsection I will show explicitly how one of Deutsch's three-bit
gates may be realized by a set of one- and two-bit gates; the result is
summarized graphically in Fig.~\ref{qcomp02}.  Here I will work only with
one version of the Deutsch gate, ``${\bf U}_\lambda$", leaving the proof for
the general Deutsch gate to the next sub-section.  ${\bf U}_\lambda$
denotes the S-matrix
\renewcommand{\arraystretch}{0.65}
\begin{equation}
{\bf U}_\lambda=\ \left(\begin{array}{cccccccc}1&\ &\ &\ &\ &\ &\ &\ \\
\ &1 &\ &\ &\ &\ &\ &\ \\ \ &\ &1 &\ &\ &\ &\ &\ \\ \ &\ &\ &1 &\ &\ &\ &\ \\
\ &\ &\ &\ &1 &\ &\ &\ \\ \ &\ &\ &\ &\ &1 &\ &\ \\
\ &\ &\ &\ &\ &\ &\cos\lambda &i\sin\lambda \\
\ &\ &\ &\ &\ &\ &i\sin\lambda &\cos\lambda \end{array}\right)\label{four},
\end{equation}
where now we have exhibited the S-matrix as an $8\times 8$ unitary matrix;
we take the basis to be the ``computational basis" labelled 0 through 7,
identified with the three-bit states $0=|0,0,0>$, $1=|0,0,1>$, ...,
$7=|1,1,1>$.  The labeling of the three bits is indicated in
Fig.~\ref{qcomp02}.
The action of ${\bf U}_\lambda$ may be expressed in words
as follows: ``Perform a rotation
of the quantum state by angle $\lambda$ in the plane in Hilbert space defined
by the state vectors $|1,1,0>$ and $|1,1,1>$ (states 6 and 7).  Leave
all other components of the state, those for which either the first of the
second bit is a zero, unchanged."  ${\bf U}_\lambda$ is obtained by four
applications of ${\bf S_Q}$ in Eq. (\ref{one}), with the identification
$\lambda=-2\pi\alpha$.

In Fig.~\ref{qcomp02} I introduce new gates {\bf X} and {\bf V}, which
operate only on pairs of bits at a time.  In a two-bit basis they have
the S-matrices
\begin{equation}
{\bf X}^{(2)}=\ \left(\begin{array}{cccc}e^{i\phi}&\ &\ &\ \\ \ &1 &\ &\ \\
\ &\ &1 &\ \\ \ &\ &\ &1\end{array}\right),
{\bf V}^{(2)}=\ \left(\begin{array}{cccc}
1 &\ &\ &\ \\ \ &1 &\ &\ \\ \ &\ &\cos\phi&\sin\phi\\ \ &\ &-\sin\phi&\cos\phi
\end{array}\right)\label{five}.
\end{equation}
The S-matrices of these gates operating in the basis of all three bits
is a direct product, e.g., ${\bf V}^{(2)}\otimes\openone$, and so is a
block-diagonal $8\times 8$ matrix, although with an ordering of rows and
columns which is determined by which pair of bits {\bf V} operates
upon.  For example, when {\bf V} operates on bits 1 and 3 as in
Fig.~\ref{qcomp02}, the full S-matrix is
\begin{equation}
{\bf V_{13}}=\ \left(\begin{array}{cccccccc}
1 &\ &\ &\ &\ &\ &\ &\ \\ \ &1 &\ &\ &\ &\ &\ &\ \\ \ &\ &1 &\ &\ &\ &\ &\ \\
\ &\ &\ &1 &\ &\ &\ &\ \\ \ &\ &\ &\ &\cos\phi&\sin\phi&\ &\ \\
\ &\ &\ &\ &-\sin\phi&\cos\phi&\ &\ \\ \ &\ &\ &\ &\ &\ &\cos\phi&\sin\phi\\
\ &\ &\ &\ &\ &\ &-\sin\phi&\cos\phi\end{array}\right)\label{six},
\end{equation}
and similarly for ${\bf X_{23}}$.  The operator {\bf N} is simply the
classical ``NOT", i.e., Eq.(\ref{two}) with $\alpha=1$.  Now, it is a
straightforward algebraic exercise, involving the multiplication of
a succession of $8\times 8$ matrices,
to show that the equation of Fig.~\ref{qcomp02}
is true, (or, in mathematical symbols,
\begin{equation}
{\bf U_\lambda}(\lambda=\delta)\simeq{\bf N_2}{\bf V_{13}}(\phi=\sqrt\delta)
{\bf X_{23}}(\phi=-\sqrt\delta){\bf V_{13}}(\phi=-\sqrt\delta)
{\bf X_{23}}(\phi=\sqrt\delta){\bf N_2}\ ),\label{seven}
\end{equation}
to first order in
the small parameter $\delta$.  To obtain ${\bf U}_\lambda$ for
any $\lambda$ to a desired degree of accuracy, it is only necessary to
concatenate a set of small rotations, by writing ${\bf U}_\lambda=
({\bf U}_{\lambda/N})^N$; the error made by using the set of two-bit
operations in Eq.(\ref{seven}) can be shown to be of order
$1/\sqrt{N}$.

\subsection{Completion of the proof: generating the entire Lie algebra}
\label{sec:level4}

The foregoing does not quite complete the proof of the universality of
two-bit gates, because I have only shown that one particular three-bit
gate (${\bf U}_\lambda$) is obtainable; Deutsch uses three others
(which he called ${\bf V}_\lambda$, ${\bf W}_\lambda$ and ${\bf X}_\lambda$)
to generate an arbitrary quantum network.  Rather than continue on in the
same pedestrian fashion for these other three cases (which gets a bit more
involved), I will show that the above results, and the
other ones which are required, may be obtained very compactly within the
language of Lie groups\cite{MWtal}.

Expressed in group-theoretic language, all the computational gates discussed
above are elements of the Lie group $U(2^3)$, and the question of universality
is the same as the question of whether the set of transformations I have
defined suffice to {\em generate} $U(2^3)$.  Deutsch has already demonstrated
that the set of $U(2^3)$ elements ${\bf U}_\lambda$---${\bf X}_\lambda$ in
turn suffice to generate the group $U(2^k)$ for an arbitrary number of bits
$k$.

The standard concept from Lie-group theory of {\em infinitesimal generators}
fits hand-in-glove with the construction of unitary logical gates.  The
infinitesimal generators {\bf H} of the Lie group are defined by
\begin{equation}
\delta {\bf U}={\bf \openone}+i\epsilon{\bf H}.\label{eight}
\end{equation}
In our problem $\delta {\bf U}$ are $8\times 8$ unitary matrices differing
infinitesimally from the identity, $\epsilon$ is an arbitrarily small
number, and {\bf H}, the generators, are $8\times 8$ Hermitian matrices.
There are 64 distinct $8\times 8$ Hermitian matrices; for later reference
I write out here a convenient set of them ${\bf H}_{\alpha\alpha}$,
${\bf H}^r_{\alpha\beta}$ and ${\bf H}^i_{\alpha\beta}$ ($0\leq\alpha<\beta
\leq 7$); their matrix elements are
\begin{mathletters}
\label{generallabel}
\begin{equation}
({\bf H}_{\alpha\alpha})_{ij}=\delta_{i\alpha}\delta_{j\alpha},\label{math1}
\end{equation}
\begin{equation}
({\bf H}^r_{\alpha\beta})_{ij}=\delta_{i\alpha}\delta_{j\beta}
+\delta_{i\beta}\delta_{j\alpha},\label{math2}
\end{equation}
\begin{equation}
({\bf H}^i_{\alpha\beta})_{ij}=-i\delta_{i\alpha}\delta_{j\beta}
+i\delta_{i\beta}\delta_{j\alpha}.\label{math3}
\end{equation}
\end{mathletters}
A key theorem of Lie-group theory is that, if ${\bf H}_1$ and ${\bf H}_2$
are generators of the group, then other generators may be obtained by
{\em commutation}, producing the {\em Lie algebra}:
\begin{equation}
{\bf H}_3=i[{\bf H}_1,{\bf H}_2].\label{comm}
\end{equation}
Moreover, one can write down an explicit expression for how the unitary
operation $\exp(i\epsilon{\bf H}_3)$ is obtained from
$\exp(i\epsilon{\bf H}_1)$ and $\exp(i\epsilon{\bf H}_2)$:
\begin{equation}
e^{i\delta(i[{\bf H}_1,{\bf H}_2])}\simeq e^{i\sqrt\delta{\bf H}_2}
e^{-i\sqrt\delta{\bf H}_1}e^{-i\sqrt\delta{\bf H}_2}e^{i\sqrt\delta{\bf H}_1},
\label{crux}
\end{equation}
this valid for small parameter $\delta$.  Thus we see that the sequence
of gates illustrated in Fig. \ref{qcomp02}
and in Eq. (\ref{seven}) is nothing more than the
execution of a commutator of the Lie algebra.

With this machinery, the question of whether two-bit gates suffice to
produce all possible three-bit unitary operations boils down to the
question of whether the successive commutation of the Hermitian generators
of our set of two-bit gates fills out the entire 64-dimensional Lie
algebra spanned by Eqs. (\ref{generallabel}).  Actually the exercise is
simpler than this, because as Deutsch showed, obtaining the generators
corresponding to just four unitary operators ${\bf U}_\lambda$,
${\bf V}_\lambda$, ${\bf W}_\lambda$ and ${\bf X}_\lambda$ suffices to
produce all of $U(8)$.  The four corresponding Hermitian generators are
${\bf H}_{66}$, ${\bf H}_{77}$, ${\bf H}^r_{67}$ and ${\bf H}^i_{67}$.
So, I forthwith show the explicit commutator expressions for these four
generators, keeping in mind that I am also allowed to introduce the
one-bit ``NOT" operation, in addition to the two-bit operations:
\begin{mathletters}
\label{gencom}
\begin{equation}
{\bf H}^r_{67}={\bf N}_2(i[{\bf H_{X_{23}}},{\bf H_{V_{13}}}]){\bf N}_2,
\label{com1}
\end{equation}
\begin{equation}
{\bf H}^i_{67}={\bf N}_2(i[{\bf H_{X_{23}}},{\bf H_{U_{13}}}]){\bf N}_2,
\label{com2}
\end{equation}
\begin{equation}
{\bf H}_{66}={\bf N}_2(-\frac{i}{4}[[{\bf H_{X_{23}}},{\bf H_{V_{13}}}],
[{\bf H_{X_{23}}},{\bf H_{U_{13}}}]]+\frac{1}{2}{\bf N}_1{\bf H_{X_{12}}}
{\bf N}_1){\bf N}_2,
\label{com3}
\end{equation}
\begin{equation}
{\bf H}_{77}={\bf N}_2(\frac{i}{4}[[{\bf H_{X_{23}}},{\bf H_{V_{13}}}],
[{\bf H_{X_{23}}},{\bf H_{U_{13}}}]]+\frac{1}{2}{\bf N}_1{\bf H_{X_{12}}}
{\bf N}_1){\bf N}_2.
\label{com4}
\end{equation}
\end{mathletters}
Here ${\bf U_{13}}$ is a two-bit gate not previously introduced; it is
similar to ${\bf V_{13}}$, having the two-bit S-matrix (cf. Eq. (\ref{five})):
\begin{equation}
{\bf U}^{(2)}=\ \left(\begin{array}{cccc}
1 &\ &\ &\ \\ \ &1 &\ &\ \\ \ &\ &\cos\phi&i\sin\phi\\ \ &\ &i\sin
\phi&\cos\phi
\end{array}\right)\label{newgat}.
\end{equation}
The Hermitian matrices ${\bf H_{V_{13}}}$, etc. are the generators
corresponding
to the designated two-bit operations, which may be obtained from a
Taylor-series
expansion of the corresponding $8\times 8$ S-matrices (e.g.,
Eq. (\ref{six})).

Eqs. (\ref{gencom}) complete the proof that all necessary three-bit
operations
can be executed using two-bit gates; the explicit sequence of gates can be
read off the equations.  This is not to say that they provide a very
practical
implementation of quantum logic.  For one thing, Eqs. (\ref{gencom}) only
provide a way of getting unitary operations with small rotation angles.
Secondly, these equations specify a rather lengthy sequence of two-bit
gates, especially Eqs. (\ref{com3}) and (\ref{com4}), for which the
analog of Fig. \ref{qcomp02} would contain a sequence of 21 gates!
Clearly it would be worthwhile to seek for more efficient techniques
for implementing some quantum computations of interest, such as the
Fourier transform of Shor.

\section{Conclusions}
\label{conc}

It appears that very rapid progress is now being made on the fundamentals
of quantum computing.  It is well to keep in mind, though, that many basic
issues of the realization of quantum computers remain unsolved or very
difficult.  The physical difficulties go well beyond the
necessity for long phase-coherence times emphasized in Sec. \ref{sec:vvd}.
As Landauer has discussed\cite{Landaue}, quantum computers suffer from
instabilities in their time evolution which are inherent to any Hamiltonian
system; in addition quantum computers cannot be error-corrected in any
traditional sense, since error correction is intrinsically
dissipative.  Considerable ingenuity will be needed if these obstacles
are to be overcome; the quantum factoring algorithm of Shor shows that
there is considerable value in overcoming these obstacles.

Another cautionary note, though, has been sounded by the computer
scientists.
There is evidence that quantum computers, while
undoubtedly more powerful than classical computers, may not be able to
solve efficiently all the famous hard problems
known to computational theorists.
For example, there is evidence that quantum computers cannot solve the
NP class of problems\cite{CHB} any more rapidly than classical computers,
although the method of proof used, employing
a so called ``random oracle", is known not to be definitive.  Indeed,
quantum mechanics has clearly created a whole new challenging and
interesting area of investigation for computational complexity theorists.
The main definite result for the time being is Shor's factoring algorithm,
which gives the hope that closely related problems such as graph
isomorphisms\cite{Shorpriv} might also have a rapid solution.

The present work shows that all quantum logic can in principle be
designed with two-bit gates; however, it does not offer any practical
design principles for quantum logic, and this remains an important
open issue for the future.  For the specific case of the Shor algorithm,
Coppersmith\cite{Copper} has very cleverly shown how both the
essentially quantum-mechanical parts of his algorithm, and the
``conventional" reversible part, may be very efficiently designed in
two-bit gates.

I wish to close by pointing out the path for new physics experiments
which is suggested by the gearbox quantum computer.  The present
proposal envisions a very ambitious program in which perhaps thousands
of quantum-mechanical operations are carried out to execute a quantum
algorithm; but even the execution of a few of the unitary operations of
Eq. (\ref{tord}) would constitute new and interesting physics.  For example,
with just one such operation a so-called
``Einstein-Podolsky-Rosen" pair\cite{Perez}
can be formed.  By spatially separating
this pair and performing single-spin measurements on the two, one would
observe the space-like non-locality unique to quantum mechanics, and
learn crucial information about dephasing times for pairs of spins.
Other unique quantum phenomena like ``teleportation"\cite{Tel} could
also be investigated.  Such investigations could well be as exciting
as the creation of the quantum computer itself, and they certainly
lie along the path to it.

\acknowledgments

I am grateful to my colleagues N. Amer, C. H. Bennett, D. Coppersmith,
N. Gershenfeld,
A. D. Kent, R. Landauer and S. Lloyd for many helpful discussions about this
work.  Thanks particularly to D. Coppersmith for supplying a crucial
piece of help in establishing the sufficiency of two-bit gates, and
for some useful references.

\begin{figure}
\caption{
The gearbox quantum computer.  The two meshed gears operate
classically, turning in synchrony.  A single quantum spin-1/2 degree of
freedom, discussed as a proton nuclear spin in the text, is firmly
attached to the end of the tips of the left-hand gear, and to the base
of the grooves of the right-hand gear.  Other gears may be added for
I/O, memory, etc.  The teeth of the left-hand gear are shown in the
shape of atomic-force microscope (AFM) tips, suggestive of the fact that
atomic spatial resolution will be necessary in the meshing of the two
gears, in order that the two spins may be brought into atomic contact.
The gears are shown with 16 and 15 teeth respectively; by making these
numbers relatively prime, it is assured that each pair of spins from the
two separate gears
may be brought into contact by turning the gearbox.}
\label{qcomp01}
\end{figure}

\begin{figure}
\caption{
Explicit demonstration of the equivalence of one of Deutsch's
three-bit gates with a sequence of two-bit gates, for infinitesimal
values of the rotation parameter $\delta$.  The S-matrices of gates ${\bf U}_
\lambda$, ${\bf X}$ and ${\bf V}$ are
described in Eqs. (\protect\ref{four},\protect\ref{five},\protect\ref{six}).
The labeling of the three bits discussed in the text is indicated,
including the primed notation for their output states.  The sequence
of two-bit gates
shown amounts to the execution of a
commutator of the generators of the $U(8)$
Lie algebra, as discussed in Sec. \protect\ref{sec:level4}.}
\label{qcomp02}
\end{figure}


\begin{references}
\bibitem{foot1} This discussion indicated that hopefully, an
ongoing goal of solid state and low temperature physics should
be to find new quantum subsystems with long $t_\phi$'s.
\end{references}
\end{document}